RESEARCH ARTICLE

# Efficiency Gains in Social Science Research: A Prospective Silicon Sampling Protocol for Pre-Treatment and Code Optimization

Author 1: Noaïm Astitou[1]

Author 2: Etienne Aucouturier[2]*

*Corresponding author: etienne.aucouturier@cea.fr

**Abbreviations:** LLM: Large Language Model; FLI: France Life Imaging; TPI: Thematic Predictability Index; FDR: False Discovery Rate; WRS: Word Recurrence Score; PCA: Principal Component Analysis

## Abstract

This study evaluates the methodological potential of using large language models (LLMs) to generate synthetic datasets prior to real-world data collection, thereby streamlining subsequent processing and statistical analysis. The approach is applied to a survey conducted within the France Life Imaging (FLI) network, leveraging three distinct LLMs (Sonnet 4.6, GPT 5.3, and Kimi 2.6) to simulate responses and prepare data processing workflows. The findings demonstrate significant time savings in code development (85–90% reuse rate) and highlight the conditional validity of this preprocessing method, which depends on the thematic predictability of the models. While synthetic data effectively captures collective perceptions, it struggles with specialized or contextualized practices, reflecting limitations documented in the silicon sampling literature, such as homogenization and reduced variance. This work extends the Phase 0 concept (Trojacka et al. 2025) from instrument validation to a full-fledged methodological preprocessing step, paving the way for scalable applications in survey research.

## Policy Significance Statement

This study shows how artificial intelligence (AI) can help policymakers and social scientists save time and resources when designing surveys. By using AI to generate synthetic data before collecting real responses, teams can pre-test questionnaires, optimize data processing, and identify potential biases—without waiting for actual survey results. This approach is especially useful for policy evaluations where speed and cost-efficiency matter. However, AI models work best for general trends (e.g., collective perceptions) but struggle with specialized or context-dependent topics (e.g., technical practices). By extending the Phase 0 concept, this method offers a practical way to improve policy research while acknowledging its limits.

[1] Commissariat à l’énergie atomique et aux énergies alternatives (CEA), Gif-sur-Yvette, FRANCE
[2] Commissariat à l’énergie atomique et aux énergies alternatives (CEA), Gif-sur-Yvette, FRANCE

## 1. Introduction

This study aims to assess the methodological value of an under-explored approach in the literature: using large language models (LLMs) to generate synthetic datasets prior to real-world data collection, thereby facilitating the preparation of data processing and statistical analysis. As an application context, we rely on a survey conducted within France Life Imaging (FLI), a national French biomedical imaging network, using a social psychology approach to examine its members' perceptions of the network's impact on their work and beliefs. We carried out this study using an anonymous online questionnaire, administered through LimeSurvey between March and May 2026, which combined closed-ended questions, Likert scales, and open-ended questions. It covered the respondents' socio-demographic profiles as well as their perceptions of the network's contributions, its impact on innovation and training, and the compliance of its practices with various ethical frameworks. The research areas of the FLI network, whose members we surveyed, are organized into four complementary domains: Molecular Imaging Agents (RE1), Technological Innovations and Instrumentation (RE2), Interventional Imaging (RE3), and Multimodal Imaging Processing and Analysis (RE4).

This work falls within the field of computer science applied to policy research, leveraging the silicon sampling technique (Argyle et al. 2023), which involves using a language model to generate a virtual population of respondents whose response distributions aim to replicate those of human groups. However, the vast majority of studies using this framework (Argyle et al. 2023; Bisbee et al. 2024; Li et al. 2025; Santurkar et al. 2023) adopt a retrospective approach: model outputs are compared to previously collected human data, following an ex post evaluation or validation logic.

Our approach differs in both its timing and purpose. The synthetic databases were generated prior to real-world data collection, with the aim of preparing subsequent processing and statistical analysis steps. This prospective approach draws inspiration from the *Phase 0* concept (Trojacka *et al.* 2025).

In this work, we examine the methodological value of this approach through two complementary questions:

Can generating silicon samples prior to real-world data collection effectively prepare for the processing and analysis of upcoming data—by structuring the code, anticipating methodological choices, and early detection of potential questionnaire design biases?

And, as a second but inseparable aspect, we ask: Is the algorithmic fidelity of the synthetically generated data sufficient to ensure that this pre-processing remains valid once real-world data become available? Or do the algorithmic fidelity gaps documented in the literature (Bisbee *et al.* 2024; Li *et al.* 2025) invalidate all or part of the preparatory work thus conducted?

To address these two questions, we use three language models with distinct characteristics (Sonnet 4.6, GPT 5.3, Kimi 2.6), whose comparison with the real-world data from the FLI survey provides the empirical material necessary for this dual evaluation. The comparison between these models therefore serves as evidence for our argument, rather than being the primary subject of the paper.

The first of these two aspects—the effectiveness of pre-processing—is assessed through a retrospective observation of the time spent developing the data processing code, rather than via a controlled experimental protocol comparing a condition with and a condition without prior use of synthetic data. This limitation, discussed in Section 5, calls for caution in interpreting the causal significance of the observed methodological benefit on this specific point. The second aspect, concerning the algorithmic fidelity of synthetic data, relies instead on standard statistical tests applied to the collected data, whose validity is not affected by this same limitation.

To implement this work, we designed a prompt describing a fictional network that replicates the characteristics of the studied network (FLI). This prompt was then submitted to the three models to generate member profiles consistent with the specified characteristics.

We then defined a standardized response format and asked these models to simulate responses to a questionnaire identical to the one administered in the actual survey. The synthetic databases thus produced were used to prepare the processing and analysis procedures for the real-world database, with the goal of optimizing the time dedicated to its exploitation once it became available.

We designed this work as a proof of concept, in the form of an exploratory case study focused on a single network and a relatively modest sample size (n = 129, representing approximately 20% of the listed members). Its aim is neither to establish general statistical validity nor to generalize numerical results to other networks, but rather to demonstrate the operational feasibility of the protocol and to precisely document the conditions for its success and failure within a well-defined context, paving the way for large-scale replication.

The three LLMs used—Sonnet 4.6 (Claude AI), GPT 5.3 (ChatGPT), and Kimi 2.6 (Kimi AI)— were selected based on their respective functionalities, deemed suitable for the task at hand, as well as their market prominence at the time of generating the synthetic databases. A fourth mixed database, referred to as the '*Mixte*' dataset, was created by merging the three synthetic databases to assess whether combining their predictions mitigates the biases inherent to each individual model. The results reported in this work are specific to these versions and the study's timeline, a limitation further discussed in the conclusion.

To enable a thorough understanding of how the synthetic databases were generated, we provide in the appendix the full set of prompts submitted to the different LLMs (in their original French version and English translation). The appendix includes the description of the fictional network used for the simulations, the requested number of observations, the instructions prompting each model to independently reproduce the network's archetypes based on this description, the studied themes, the complete set of questions submitted to them, as well as the various significance tests performed. Making these elements available aims to ensure the transparency of the methodological approach and to facilitate the reproducibility of the experiment.

## 2. Definitions and State of the Art

### 2.1. Silicon sampling

The term *silicon sample* originates from the foundational article by Argyle, Busby, Fulda, Gubler, Rytting, and Wingate, published in 2023 in *Political Analysis* (31(3), 337-351) under the title « Out of One, Many: Using Language Models to Simulate Human Samples ». Silicon samples refer to synthetically generated participants designed to mimic human respondents in order to describe, explain, and predict human behaviors, typically calibrated to predefined target groups and demographic segments. In their work, the authors create silicon samples by fine-tuning GPT-3 on thousands of sociodemographic biographies derived from real participants in several major American surveys. This contribution introduces a second, inseparable concept: *algorithmic fidelity*. It refers to the ability of a properly instructed LLM to generate a virtual population of respondents whose response distributions replicate those of existing human groups. The authors demonstrate in particular that the model's bias is not uniform but rather nuanced and correlated with the demographic characteristics provided beforehand. This work, considered foundational in the field, is now applied across various disciplines, including marketing (Sarstedt *et al.* 2024) and psychology (Dillion *et al.* 2023).

A parallel and complementary framework is proposed by Horton, Filippas, and Manning in a National Bureau of Economic Research report: that of *homo silicus* (2023). Their central thesis is that LLMs constitute implicit computational models of humans, which researchers can leverage by endowing the model with preferences and constraints, then observing its behavior in simulated scenarios. This framework is directly relevant to our protocol. While Argyle et al.(2023) validate the ability of LLMs to replicate opinion distributions through demographic conditioning, Horton et al. (2023) validate their capacity to behave as agents with preferences—a logic we employ when we instruct the AIs to construct archetypes of members of the fictional network and then have them respond to the questionnaire. Our protocol combines the two foundational contributions: generating synthetic profiles consistent with the described network (following the *homo silicus* logic) and simulating responses to a survey identical to the one administered to the network's members (following the silicon sampling logic).

However, recent literature significantly tempers the enthusiasm surrounding silicon sampling by highlighting three limitations that can be directly applied to our own results.

First, Bisbee et al. (2024) demonstrate that *silicon sampling* overfits to majority opinions and struggles to faithfully represent marginal positions. Second, Li et al. (2025) identify two complementary shortcomings: a structural coherence breakdown across different levels of aggregation, as well as a homogenization effect resulting in the underrepresentation of minority opinions. Third, Santurkar et al. (2023) reveal a substantial misalignment between the views reflected by the models and those of real demographic groups, a misalignment that persists even after explicit conditioning. These studies share a common feature: they compare LLM outputs to previously collected human data, following a retrospective evaluative or critical logic. Our protocol differs in both its purpose and timeline: the synthetic databases were generated prior to obtaining the survey results, with the methodological aim of preprocessing the actual data.

### 2.2. Synthetic Data Generation for Research

Grossmann et al. (2023) show that LLMs offer opportunities to test theories of human behavior at scale, while posing a central methodological challenge of ensuring transparent and reproducible research. Overton et al. (2025) , on the other hand, synthesize the main categories of LLM applications in research: simulating responses in experiments and surveys, augmenting existing datasets, testing research instruments, and generating exploratory data when traditional collection is impracticable.

A distinct methodological use involves completing partially collected data via LLMs, rather than generating it entirely. Kim and Lee (2026) thus developed a language model fine-tuned on the American General Social Survey (1972–2021) for three tasks: imputing missing data (AUC = 0.87), retrodicting missing opinions (AUC = 0.86), and predicting opinions never previously surveyed (AUC = 0.70, with more modest performance).

These various uses allow us to position our protocol at the most ambitious end of this methodological continuum, as it does not aim to complete or reconstruct existing data but rather to fully generate a synthetic database from a simulated environment.

A recent development involves further personalizing model conditioning by relying on in-depth qualitative interviews rather than simple demographic characteristics. In this context, Park et al. (2024) propose a generative agent architecture representing 1,052 real individuals based on semi-structured interviews and questionnaires. The resulting agents achieve up to 86% of the test-retest consistency level observed among *General Social Survey* participants, illustrating the ability of LLMs to simulate individual attitudes and behaviors without specific training on the considered questions.

Finally, Madden (2025) recommends treating synthetic agents as high-capacity predictors under specified conditioning, rather than as sources of calibrated probabilistic belief, and thus actively measuring their calibration by subgroup rather than assuming it, as algorithmic fidelity is conditional and demographically structured.

This observation aligns with earlier findings: Aher et al. (2023) , in testing models' ability to replicate classic experiments in economics and psychology, document a phenomenon they term *hyper-accuracy distortion*—where generated responses cluster too tightly around the expected answer, erasing some of the variability inherent in real human responses. Bisbee et al. (2024) document a related phenomenon: a lower diversity in LLM-generated responses compared to human survey data.

### 2.3. Using Silicon Sampling Prior to Real-World Data Collection

Wang et al. (2026) present a five-step framework for integrating LLMs into survey research, with a pilot testing phase (traditionally conducted with real respondents) now replaced by simulated responses conditioned on demographic profiles. The authors explicitly clarify that the goal is to support design, not to fabricate data: synthetic responses inform questionnaire and analysis plan decisions prior to fieldwork, without being included in the final analytical dataset.

Trojacka et al. (2025), for their part, present the work most closely aligned with our approach, both methodologically and in terms of application. In the biomedical field, they propose an *in silico* pre-validation framework using LLMs to stress-test an instrument prior to its clinical deployment, by generating a synthetic cohort of 500 patient profiles with detailed clinical attributes. The authors designate this step as a *Phase 0* of instrument development.

Valenzuela et al. (2025) place this practice more broadly within the survey lifecycle, distinguishing three phases of LLM intervention—questionnaire development, synthetic data generation, and open-ended response analysis—and note that these models can detect certain types of design errors even before traditional pilot studies.

This body of work confirms that using synthetic data prior to real-world collection is a developing practice, but that existing studies focus on a specific and relatively narrow objective: quality control of the measurement instrument itself.

### 2.4. Our hypotheses

Our approach lies at the intersection of three areas that the literature has thus far treated separately. First, following the work of Wang et al. (2026) and Trojacka et al. (2025), we use synthetic generation as a tool for pre-validating a questionnaire prior to its deployment. Second, by directly comparing the distributions of synthetic responses to the observed distributions after actual data collection, we assess the algorithmic fidelity of the models, building upon silicon sampling and the critical work of Bisbee et al. (2024) and Li et al. (2025), but through a prospective rather than retrospective protocol. Third, we explore a still under-documented use: leveraging synthetic data to prepare the statistical processing of data that have not yet been collected, enabling us to anticipate analyses, organize the associated code for data processing, and reduce the time required for exploiting real-world data.

This final dimension extends the concept of Phase 0, introduced by Trojacka et al. (2025) in the biomedical field to denote a pre-validation step conducted in silico for an instrument prior to its deployment. While the literature primarily employs this approach as a quality control tool for questionnaires, we hypothesize that it can constitute a genuine methodological pre-processing step. The question, therefore, is no longer merely whether LLMs produce plausible responses, but whether generating synthetic data prior to collection enables the efficient preparation of upcoming data processing, and whether the algorithmic fidelity of the models is sufficient to ensure that this pre-processing remains valid once real-world data become available.

Two hypotheses directly follow from this:

H1, a tendency for models to overrepresent majority positions while underrepresenting marginal ones (Bisbee *et al.* 2024); and H2, lower variance in synthetic datasets compared to real-world datasets (Aher *et al.* 2023; Bisbee *et al.* 2024).

We first present the protocol for generating synthetic datasets and the methods used to compare them with real-world data—distributional analyses, Thematic Predictability Index, Word Recurrence Score, and clustering analysis.

We then examine the results obtained along these four dimensions: the models' ability to reproduce the sociodemographic profiles of the network, their thematic accuracy, the lexical fidelity of open-ended responses compared to real-world data, and the internal structure of the data.

Finally, we document the methodological gains associated with this prospective approach and its conditional validity in light of the observed fidelity gaps, before returning, in the conclusion, to the two complementary questions raised in the introduction.

## 3. Method

### 3.1. Protocol for Generating Synthetic Datasets

A series of prompts (provided in the appendix) describing a fictional network that replicates the structural characteristics of the France Life Imaging (FLI) network was submitted to three language models: Sonnet 4.6 (Claude AI), GPT 5.3 (ChatGPT), and Kimi 2.6 (Kimi AI). These models were selected based on their suitability for the task and their market prominence. In the first stage, each model was tasked with constructing, based solely on this description, a mapping of plausible member profiles and archetypes for the network. In the second stage, the actual questionnaire administered to FLI members was submitted to each model, with the instruction to simulate responses from fictional members consistent with the previously generated profiles. Each model was prompted to generate responses for 150 participants (representing approximately a 20% response rate of the surveyed population).

The expected response format was explicitly defined for each question, replicating the conditions of a questionnaire administered via an online survey platform (LimeSurvey). This structuring ensured the compatibility of the synthetic datasets with the real-world dataset, a necessary condition for the subsequent reuse of the code associated with data processing (see Section 5). A fourth dataset, referred to as the '*Mixte*' dataset, was created by concatenating the three synthetic datasets (Claude, Kimi, ChatGPT), followed by a random shuffling of the observations (with a fixed seed for reproducibility) and truncation to the first 150 resulting entries.

### 3.2. Comparative Analysis Distribution

The sociodemographic distributions of the four synthetic datasets (Claude, ChatGPT, Kimi and *Mixte*) were compared to those observed in the real FLI dataset (n = 129) using tests appropriate to the nature of the variables: two-proportion comparison tests for binary variables and certain categorical variables, Chi² tests for multi-category categorical distributions, and Kolmogorov-Smirnov tests (distribution shape) and Wilcoxon tests (central tendency) for continuous variables.

The handling of multiple comparisons varies depending on the nature of the tested question. For variables with multiple responses (profession, affiliation hub, expertise network), each proportion test addresses an independent and directly interpretable question in itself (the gap between AI and FLI for a given category), without drawing a global conclusion from the set of categories tested simultaneously: these tests are therefore reported based on their raw p-values. In contrast, False Discovery Rate (FDR) correction is systematically applied to the tests used in constructing the Thematic Predictability Index (TPI), where the aggregation of multiple variables into a single composite score per theme—and subsequently into a global score—specifically exposes the analysis to a risk of cumulative error in the synthesized conclusion. FDR control is therefore necessary to ensure the validity of this aggregated conclusion.

The handling of missing values varies depending on the nature of the variable. For single-response variables (e.g., gender, year of integration), a missing value is treated as genuinely absent data and excluded from the analysis on a case-by-case basis. For variables encoded as checkboxes allowing multiple responses (e.g., decade of birth, profession, affiliation hub, expertise network), the absence of selection for a given category is instead treated as a non-selection (value = 0), consistent with the original form's structure, where an unchecked box corresponds to a negative response rather than a non-response.

### 3.3. Thematic Predictability Index (TPI)

To synthesize the models' ability to reproduce observed distributions, a Thematic Predictability Index (TPI) was constructed thematically. For binary variables, the absolute difference between the positive proportion observed in each synthetic database and that observed in FLI is calculated; for Likert-scale variables (1-5), the absolute difference between means is used. Each deviation is standardized into a score ranging from 0 (maximum divergence) to 100 (perfect predictability). The TPI for a thematic area corresponds to the average of the scores of the variables that compose it; a global score is obtained by averaging all thematic scores, excluding the 'profile' thematic. The retained interpretation distinguishes five tiers: near-perfect reproduction (≥ 90), very strong (80-89), good (70-79), moderate (60-69), and significant divergence (< 60) in predictive capability. To control the risk of error associated with multiple testing, each variable is tested individually (proportion test, Welch's t-test, or Chi² test, depending on its nature), and the p-values are then corrected using the Benjamini-Hochberg FDR (False Discovery Rate) method (Benjamini and Hochberg 1995), separately for each model.

The eight thematically coded areas are: Contribution (CON), Impact (IMP), Training (FOR), Ethics (ETH), Scientific Integrity (INT), Bioethics (BIO), Radioprotection (RAD), and Sustainability (PER).

### 3.4. Word Recurrence Score (WRS)

To evaluate the lexical proximity of open-ended responses, a second index was developed. Free-text responses are standardized (removal of accents, punctuation, and numbers; tokenization; removal of short words and stop words) and then represented by their relative frequency of occurrence in each database's corpus. The WRS combines three components:

- A Jaccard index (35%), measuring vocabulary overlap regardless of frequencies.
- Cosine similarity (40%), measuring the alignment of frequency distributions.
- An asymmetric weighted recovery index (25%), centered on the real database, measuring the coverage of the important vocabulary from the real database by the synthetic databases.

### 3.5. Cluster Analysis

An independent cluster analysis was conducted on each of the five databases, without any constraint on the number of clusters being identical across databases. Only numerical variables (binary and Likert-scale) were retained, after excluding text fields, residual variables, and variables with near-zero variance. Missing values were imputed using the median, and min-max normalization was applied to all variables. The optimal number of clusters for each database was determined by aggregating three normalized criteria: the elbow method, average silhouette width, and the Calinski-Harabasz index (Caliński and Harabasz 1974). A K-means algorithm (25 initializations, nstart = 25) was then applied. The quality of each segmentation was evaluated using the average silhouette score and explained variance, its non-triviality using the Hopkins statistic (Hopkins and Skellam 1954), and its robustness using a bootstrap stability test (Jaccard index on resampling). The results were visualized using Principal Component Analysis (PCA), and the average profiles of each cluster were characterized based on the ten most discriminative variables.

## 4. Results

### 4.1. Sociodemographic and Structural Profiles

Table 1 below summarizes the observed deviations between the four synthetic databases and the real database for sociodemographic variables.

Table 1 — Summary of sociodemographic and structural gaps, AI vs FLI

| Variable | Main gap | Closest model | Significant divergences | Statistical test |
|---|---|---|---|---|
| Gender | Overestimation of the proportion of women (gap of 6 à 19.4 pts) | ChatGPT | 1/4 models | Two-proportion comparison |
| Birth decade | Underestimation of the dominant decade (1970-1979) in favor of 1980-1989 | Claude | 4/4 models | Chi-squared (simulated p) |
| Year of integration | Median year systematically later (gap of 5 to 5 years) | Claude / Kimi / ChatGPT / Mixte | 4/4 models | Kolmogorov-Smirnov / Wilcoxon |
| Profession | Variable over/underestimation depending on profession | Claude / Kimi | 5-8/9 categories depending on the model | Two-proportion comparison (by profession) |
| Affiliation hub | Correct overall hierarchy, overestimation of some minority hubs | Claude / Mixte | 0-1/11 categories depending on the model | Two-proportion comparison (by hub) |
| Expertise network | Underestimation of RE2, overestimation of RE3/RF depending on the model | Mixte | 1-3/5 categories depending on the model | Two-proportion comparison (by network) |

"Closest model": model(s) with the smallest gap relative to FLI (several models listed in the event of a tie). The count columns ("x/9", etc.) show the number of categories tested that are significantly different from FLI (raw $p < 0.05$), across all AIs.

The overarching observation is one of overfitting of the models to majority categories, consistent with hypothesis H1: the three LLMs correctly reproduce dominant trends (decades, hubs, most represented professions) but smooth out minority positions. The only bias universally shared by all models, regardless of their individual statistical significance, concerns the temporal offset in network integration—consistent with the absence, in the description provided to the models, of any information about the pace of membership over time.

In accordance with the multiple comparison processing rule outlined in Section 3.2, the significances reported in Table 1 for multiple-response variables (profession, hub, expertise network) are based on the raw p-value of each individual test, rather than a p-value corrected using the FDR method.

### 4.2. Thematic Predictability Index (TPI)

The global IPT scores range from 73.8 (ChatGPT) to 81.3 (Claude), with Kimi (79.1) and *Mixte* (79.0) in between. It should be noted that the interpretation thresholds used (≥ 90, 80-89, 70-79, 60-69, < 60) constitute an exploratory scale intended to facilitate score interpretation and not a statistically validated boundary system; their use should be understood as a qualitative benchmark rather than a formal test. The best performances are observed for the thematic areas directly deducible from the institutional description of the network—Contribution, Impact, and Sustainability—with predictability scores ranging from 79.9 to 95.3. Conversely, questions related to the Ethics of Radioprotection represent the main limitation (scores from 50.3 to 60.9), followed by those concerning Training, two thematic areas referring to specialized professional practices and highly dependent on individual context. The thematic areas of General Ethical Issues, Bioethics, and Scientific Integrity occupy an intermediate position.

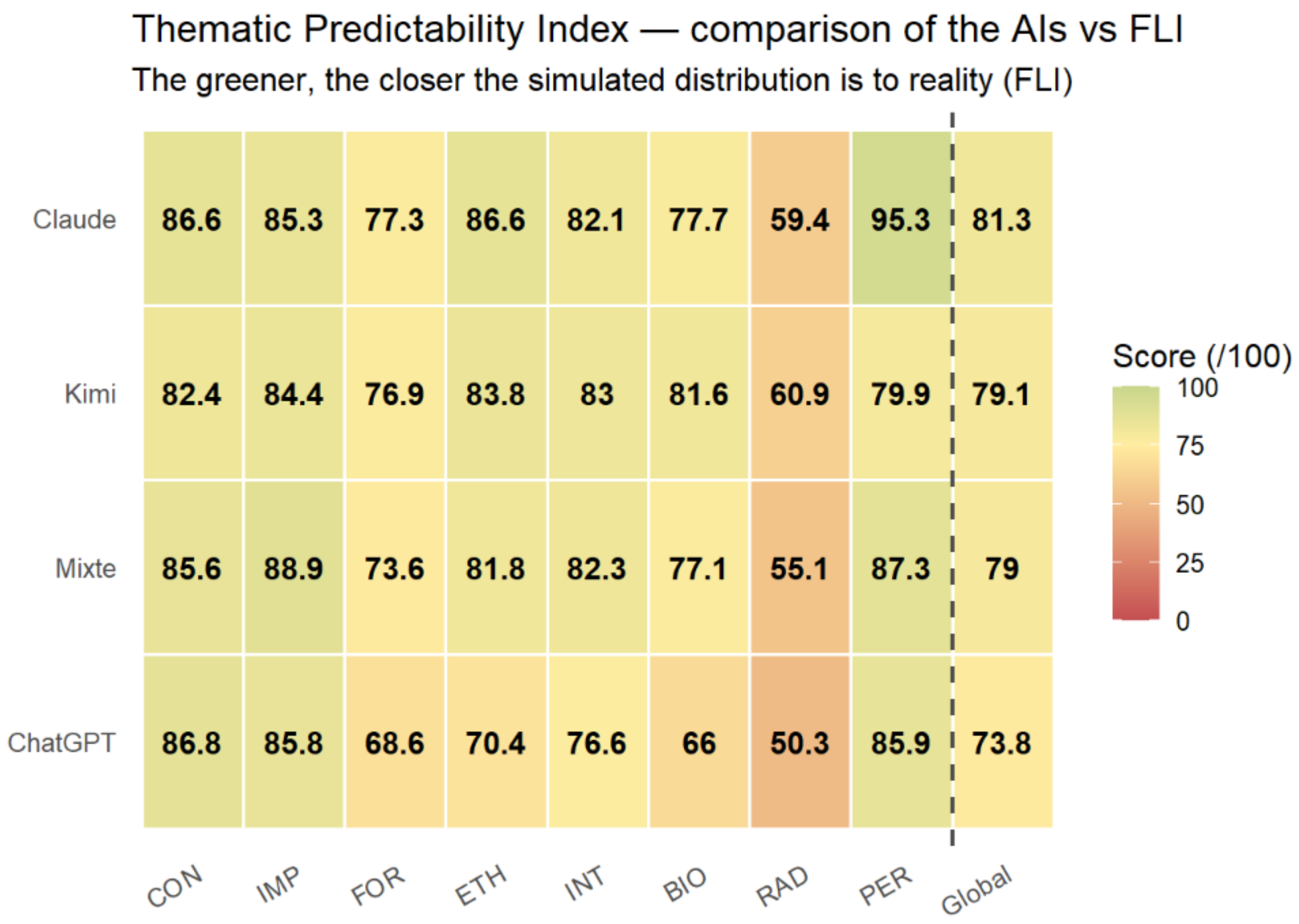

After FDR correction, the ranking based on the proportion of variables significantly different from FLI partially diverges from the ranking by average score: Kimi, although second in the global score, shows the lowest proportion of variables diverging from the real database (65.0%), followed by Claude (71.4%), *Mixte* (72.7%), and ChatGPT (84.4%).

The TPI, by aggregating standardized deviations regardless of the substantive nature of the variables, prioritizes inter-thematic comparability at the expense of weighting by the perceived importance of each variable; this methodological choice must be kept in mind when interpreting the scores.

### 4.3 Word Recurrence Score (WRS) of Open-Ended Responses

The WRS scores are uniformly low (no model reaches 50/100), revealing a structural lexical gap between generated texts and real responses. The Mixed database achieves the best relative performance (up to 23.2 on the question regarding the international impact of the network), likely due to the lexical diversity induced by its hybrid nature. This lexical smoothing—relying on generic formulations at the expense of technical and contextualized vocabulary—constitutes the most direct confirmation of the homogenization phenomena documented by Li et al. (2025) and the reduced variance documented by Bisbee et al. (2024). However, a methodological caveat applies to Kimi, as its complete lack of textual content for two variables (International Impact, National Impact) stems from a generation gap distinct from the homogenization phenomenon itself.

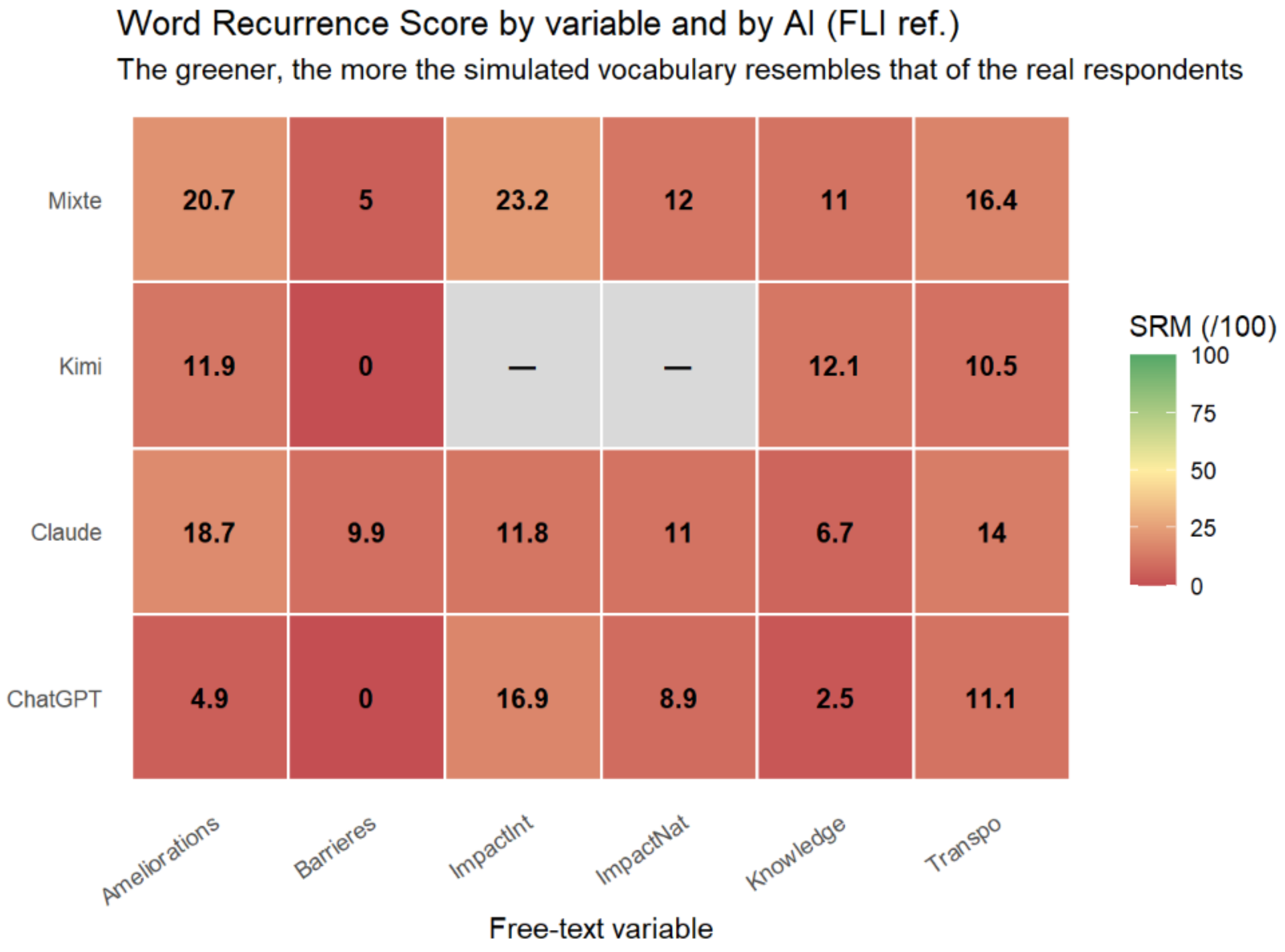

To verify the robustness of the model ranking with respect to the choice of weights (35/40/25), several alternative weighting schemes were tested (equal weighting, and inversion of weights between Jaccard and cosine similarity); the results of this sensitivity analysis, confirm the stability of the relative ranking of the models regardless of the weighting scheme used.

### 4.4 Internal Data Structure (Clustering)

Claude, *Mixte*, and FLI converge toward an optimal *k* of 2, compared to 5 for Kimi and 8 for ChatGPT. FLI exhibits the best cluster readability (14% and 7.7% of variance explained by the first two components, compared to 3-6% for synthetic databases). The bootstrap stability clearly distinguishes the segmentations: Claude (0.903), Kimi (0.909), and *Mixte* (0.836) show 100% stable partitions, while ChatGPT displays total instability (0.333), indicating over-segmentation without a robust underlying structure and no substantive interpretability.

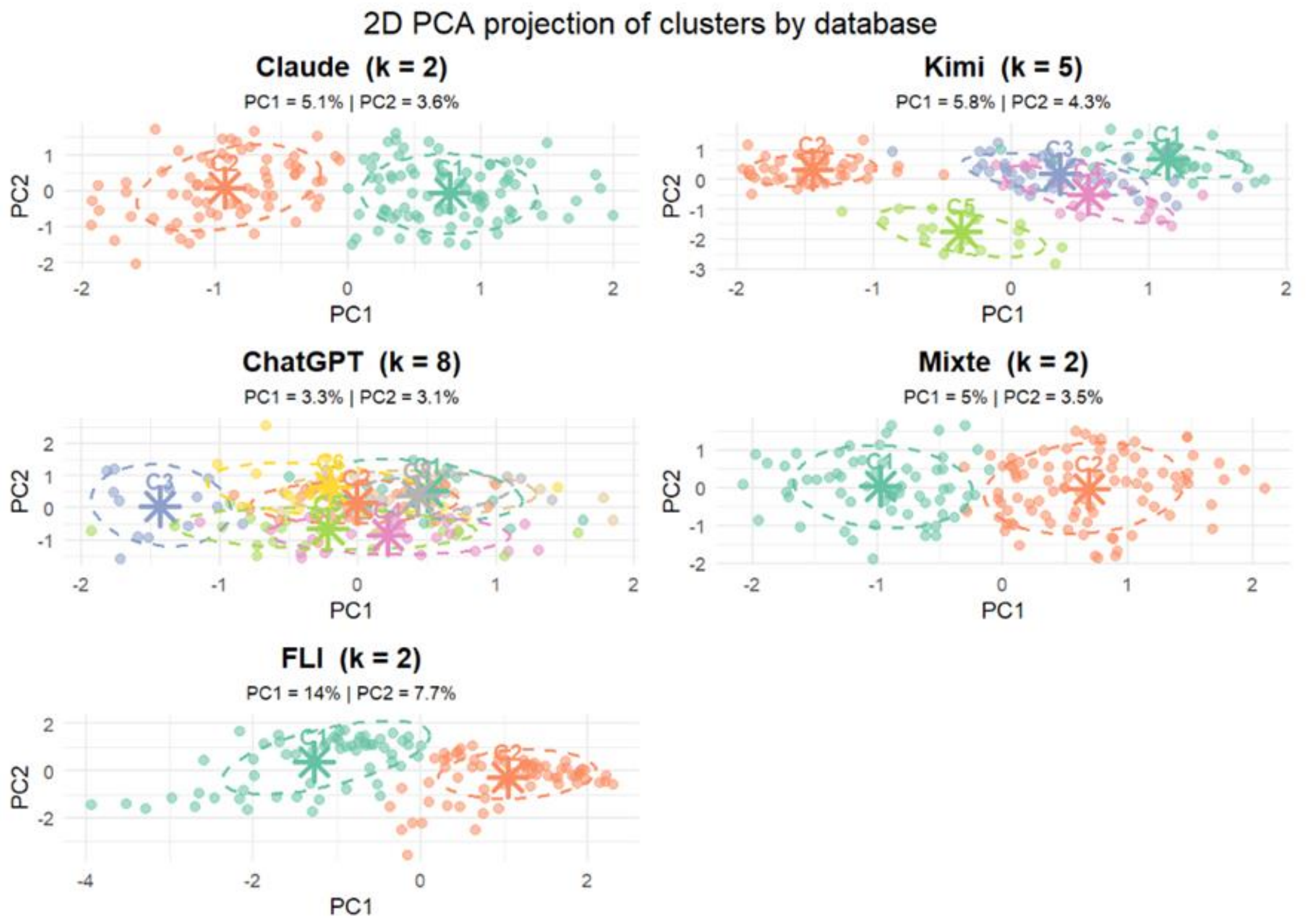

On a qualitative level, the FLI survey empirical data contrast a profile highly engaged in ethical and prevention dimensions with a profile much less marked on these same dimensions. Claude partially reproduces this structuring (academic vs. safety protocols opposition). Kimi fragments more based on professional and disciplinary criteria than perceptual ones. *Mixte* identifies differentiated clusters on ethical and structural variables, but with less clarity, likely due to its very construction: a random mix of the three synthetic databases, where the biases specific to each model tend to mutually attenuate rather than converge toward a marked structure. ChatGPT's segmentation, being unstable, does not lend itself to substantive interpretation.

## 5. Methodological Gain and Conditional Validity of Pre-processing

Beyond the statistical fidelity of synthetic databases to the real database, the primary value of this prospective approach lies in its contribution to the analytical preparation process itself. The development of the code associated with data processing spanned approximately two months, with daily but fragmented work performed on the synthetic datasets generated by AI models. This extended timeline allowed for a more methodologically favorable approach to data processing than is typically imposed: the immediate availability of data, albeit synthetic, freed this work from the time constraints usually associated with the post-collection processing phase, favoring a more deliberate reflection on the choice of statistical methods and the overall organization of the code.

After receiving the real survey data, which we publish in another social science article, the code developed beforehand required approximately one week of adjustments, primarily dedicated to adding complementary analyses that had not been anticipated during the preparatory work on the synthetic databases. This timeline contrasts favorably with the time spent on the initial design: approximately 85 to 90% of the code developed using synthetic databases could be directly reused on the real database without structural modifications. The necessary adjustments mainly involved analyses that could not be anticipated from the synthetic databases alone, rather than the architecture of the pre-developed code itself.

This time gain must, however, be interpreted in light of the upstream design effort, particularly the care taken in guiding the models during response generation. Each question in the survey was accompanied by a specific response instruction for the models, specifying the expected format, in a logic that replicates the conditions of a questionnaire administered via an online survey platform (LimeSurvey). This explicit structuring of the response format, regardless of the statistical fidelity of the responses themselves, constitutes a necessary condition for code reusability: it ensured structural compatibility between the synthetic databases and the real database, and thus the portability of the code associated with data processing from one database to another.

It should be noted that this estimation of methodological gain is based on a retrospective and qualitative observation rather than a controlled experimental design: no counterfactual condition (development of the code associated with data processing without prior use of synthetic databases) was implemented to precisely isolate the effect of the prospective approach. However, the observed contrast between the two months dedicated to the initial design and the one week required for final adaptation nevertheless constitutes converging evidence in favor of the hypothesis of significant time savings. A controlled experimental design—such as a replication of the protocol on a second network, comparing development time with and without prior use of a synthetic database—would allow for a more rigorous isolation of the specific effect of the prospective approach.

The overall time gain exhibits heterogeneity across thematic areas, consistent with the fidelity gaps documented in Section 4. This does not concern the technical functionality of the code, which remains operational, but rather the relevance of the analyses it produces, if those analyses rely on assumptions derived from the data structure. When these assumptions are close to reality, the pre-prepared analysis remains relevant; when they deviate, analytical revision may be necessary.

This heterogeneity follows the logic identified by the TPI: collective perceptions (Contribution, Impact, Sustainability) are better predicted by the models than individual or specialized practices (Radioprotection, Training). Analyses involving variables related to individual or specialized practices thus accounted for the majority of the time spent on final adjustments. The methodological gain therefore appears conditional on the nature of the variable processed: the higher the algorithmic fidelity, the less analytical revision is required.

## 6. Conclusions

The various analyses conducted in this work allow us to draw several insights, which should be interpreted in light of the methodological objective that structured our approach.

These results provide a nuanced answer to the two complementary questions posed in the introduction. Regarding the first aspect—the effectiveness of preprocessing—the documented contrast in Section 5 between the two months of development using synthetic data and the one week of adjustment required upon receiving real data constitutes converging evidence in favor of a real methodological gain, subject to the limitations inherent in a retrospective observation rather than a controlled design. Regarding the second aspect—the sufficiency of algorithmic fidelity—the results converge toward a conditional answer: sufficient for thematic areas related to collective perceptions, directly deducible from the institutional description of the network, but insufficient for thematic areas referring to individual or specialized practices, for which only real data allow for a fine-grained analysis.

These results also allow us to revisit the two hypotheses formulated in the introduction.

H1, which posits that models tend to overrepresent majority positions—of the overall active population—at the expense of marginal positions, is confirmed by the distribution of birth decades: all three models systematically underestimate the 1970-1979 decade, despite it being the most represented in the real data, in favor of an overestimation of the 1980-1989 decade, which is more consistent with the general demographic structure of the French active population.

H2, related to the lower variance of synthetic responses compared to human responses, finds clear empirical support in the WRS, which are uniformly low for the four synthetic databases, reflecting a marked lexical smoothing: the models produce more generic and less technical formulations than real respondents, even when their numerical responses are statistically close to reality.

On a thematic level, the interpretation emerging from the Thematic Predictability Index (TPI) is more nuanced than a simple opposition between collective perceptions and individual issues. The best performances concern the thematic areas most directly deducible from the institutional description of the network—Contribution, Impact, and Sustainability—for which the nature and missions of the network made certain responses statistically predictable by inference, without requiring access to real data. Conversely, the poorest performances concern questions related to the ethics of radioprotection and, to a lesser extent, training within the FLI network: these are thematic areas that refer to operational and specialized professional practices, highly dependent on the respondents' level of expertise and individual context, and thus more difficult to anticipate based solely on the general description of the network. Questions related to ethics, bioethics, and scientific integrity fall into an intermediate position: the models reproduce their broad trends but struggle to capture the nuances linked to individual trade-offs and disciplinary disparities. This gradation, rather than a simple dichotomy between the collective and the individual, outlines a hierarchy of predictability based on the extent to which a thematic area can be deduced from the institutional description of the network, as opposed to thematic areas rooted in contextual professional practices.

These results also provide insights into the limitations documented in the silicon sampling literature. The homogenization phenomenon described by Li et al. (2025) is echoed in our findings, both at the level of quantitative variables (underestimation of minority categories, unstructured over-segmentation) and in open-ended text, where the observed lexical smoothing constitutes the clearest confirmation of this mechanism in our study. The lower variance of LLM-generated responses documented by Bisbee et al. (2024) is similarly verified.

Our study also has several limitations that should be highlighted. The real sample, although substantial (129 responses, or approximately 20% of the listed network members), remains relatively modest in size for certain statistical tests, particularly those involving underrepresented subcategories, and is only applied to a single network, which limits the generalization of results to other institutional contexts. Finally, our protocol does not measure model calibration by demographic subgroup, whereas Madden (2025) emphasizes that algorithmic fidelity is conditional and demographically structured: such an evaluation would constitute a natural extension of this work.

Finally, the results obtained are specific to the versions of the models used (Sonnet 4.6, GPT 5.3, Kimi 2.6) and the date of the study. It is important to distinguish here between the reproducibility of the protocol—the method of generation, response format standardization, and statistical comparison remains applicable regardless of the model version used—and the reproducibility of the results, which is not guaranteed: TPI scores, WRS scores, and the cluster structures obtained with subsequent generations of these models may differ significantly, in either direction, as the algorithmic fidelity of LLMs evolves. This distinction limits the scope of the numerical results reported here, without calling into question the validity of the protocol itself. Periodic replication would constitute a useful extension to track this evolution over time.

These results must finally be considered in the context of the initial objective of this approach: not to perfectly reproduce the survey data, but to structure the code associated with data processing in advance. The documented methodological gain (see Section 5)—nearly two months of development on synthetic data versus one week of adaptation, with 85-90% of the code reused—confirms the relevance of this approach, provided its scope is clearly defined: it is all the more useful when questions relate to institutional perceptions deducible from the network's description, and all the more limited when they involve specialized or contextualized practices. The concept of *Phase 0* (Trojacka *et al.* 2025), initially conceived as a quality control step for an instrument, finds an extension here as a full-fledged methodological pre-processing step.

**Acknowledgments.** The authors are grateful for the administrative support provided by CEA
**Funding statement.** none
**Competing interests.** none
**Data availability statement.** The complete data from our study (the three synthetic databases, the mixed synthetic database, the FLI survey database, and the code used to analyze them) is published, at the following URL: https://doi.org/10.57745/H3YBNG
**Author contributions.** Conceptualization: N.A; E.A. Methodology: N.A; E.A. Data curation: N.A. Data visualisation: N.A Writing original draft: N.A; E.A. All authors approved the final submitted draft.